\documentstyle[prl,aps,epsf]{revtex}
\tighten
\begin{document}
\draft
\twocolumn[\hsize\textwidth\columnwidth\hsize\csname@twocolumnfalse%
\endcsname

\preprint{}

\title{Theory of Doped Excitonic Insulators}
\author{Leon Balents and Chandra M. Varma}
\address{Room 1D-368, Bell Labs, Lucent Technologies, 700 Mountain Ave, Murray
  Hill, NJ 07974}

\date{\today}
\maketitle

\begin{abstract}
The theory of doped excitonic insulators is reinvestigated in light of
recent experiments on hexaborides.  For the appropriate
valley-degenerate $X_3,X_3^\prime$ band structure, ``intra-valley''
condensation is energetically favored.  Ferromagnetism occurs upon
doping due to the quenching of kinetic
energy at the otherwise direct {\sl first-order} excitonic
insulator--metal transition.  The phase diagram
includes states of spatially inhomogeneous density and magnetization
at low temperatures.
\end{abstract}
\vspace{0.15cm}

]
\narrowtext



Recent dramatic measurements by Fisk {\sl et.
  al.}\cite{Hexaborides}\ on very lightly-doped divalent hexaborides
(${\rm La}_x{\rm Ca}_{1-x}{\rm B}_6$, ${\rm La}_x{\rm Sr}_{1-x}{\rm
  B}_6$, etc.), have revived interest in the physics of {\sl excitonic 
  insulators}, which are coherent condensates particle-hole excitations
that may occur in semi-metals with {\sl slightly} overlapping or gapped
conduction and valence bands.   The
theory of such {\sl excitonic insulators} was developed in the 1960's
and 70's (see Ref.~\onlinecite{HalperinRice}\ for a review and
original references).  Band
structure calculations predict that ${\rm CaB}_6$ and ${\rm SrB}_6$
are indeed at the border between very small gap semiconductors and
slightly overlapping electron-hole metals (with gap/overlap of $-0.08
eV < E_G < 0.08 eV$)\cite{Monnier}.  These materials are thus prime
candidates for 
an excitonic instability.  The experiments observed ferromagnetism
below $T_c \sim 600K$ with a small moment maximized
around $x=0.005$.  Early theoretical work\cite{Volkov}\ predicted
high-temperature ferromagnetism in doped EI's, and the same connection 
was recently made for the hexaborides.\cite{Zhito}\ 

\begin{figure}[htb]
\hspace{0.5in}\epsfxsize=3.25in\epsfbox{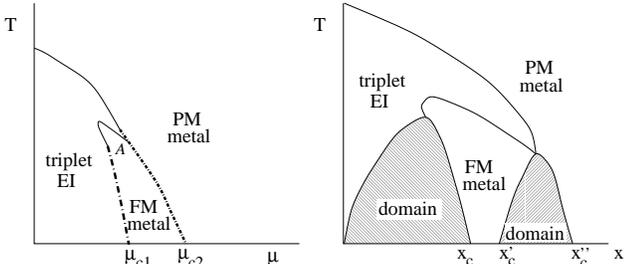}
\caption{Schematic phase diagrams in temperature versus chemical
  potential $\mu$ and doping $x$ supposing the EI is fully gapped for
  $x=0$.  The thick dashed lines indicate first-order boundaries.  If
  for $x=0$ the system is not fully gapped, a region of triplet EI 
  persists at $T=0$ for small $x$.  The order of the phase
  boundaries near the point $A$ is sensitive to the detailed
  parameters of the model.}
\end{figure}

In this letter the theory of EI's is reinvestigated, taking into
account the three-pocket band structure appropriate to the
hexaborides.  A number of significant errors in the previous work on a
two-band model are also corrected.  In particular, we find that a
first-order transition as chemical potential is varied and a
consequent jump in the equilibrium density is crucial in the physics
of ferromagnetism (see Fig.~1).

To proceed, we make the crucial physical assumption that the
characteristic energy scales of the EI are much smaller than the
bandwidth.  A natural parameter characterizing this smallness is the
ratio of the band gap $E_G$ ($E_G <0$ for overlapping bands) to the
bandwidth $W$.  For $|\Gamma| = |E_G|/W \ll 1$, the physics is
controlled by small momentum-exchange processes.  In this limit, the
long-range part of the Coulomb potential dominates, and much of the
behavior becomes insensitive to the details of the band structure.  To
make this explicit, we expand the electron field operators
$c_\alpha^{\vphantom\dagger},c_\alpha^\dagger$ ($\alpha
=\uparrow,\downarrow$ is a spin index):
\begin{equation}
  c_\alpha^\dagger({\bf r}) = \sum_{ia} \int_{\bf k}^\Lambda
  \phi^{\vphantom\dagger}_{i{\bf Q}_{a}+{\bf k}}({\bf r})
  \psi_{ia\alpha}^\dagger({\bf k}).
  \label{formalexpansion}
\end{equation}
Here $\phi_{i{\bf k}}({\bf r})$ is a Bloch function at quasi-momentum
${\bf k}$ with $i=1,2$ for valence and conduction states,
respectively.  The integration ($\int_{\bf k} \equiv \int \! {d^3{\bf
    k} \over {(2\pi)^3}}$) is restricted to spheres of radius
$\Lambda$ around the regions of band proximity, centered at the three
wavevectors ${\bf Q}_{a} =(\pi,0,0),(0,\pi,0),(0,0,\pi)$
(we measure all lengths in units of the lattice spacing).  The
``flavored'' fermion fields (Fourier transformed back to real space) obey $\{
\psi_{ia\alpha}^{\vphantom\dagger}({\bf r}),
\psi_{jb\beta}^\dagger({\bf r'}) \} = \delta_{ij}\delta_{ab}
\delta_{\alpha\beta} \delta({\bf r-r'}).$ The effective mass
approximation is valid for $\Gamma \ll 1$, since then we can take
$\Lambda \ll 1$.  Thus $H= H_0+H_C$, with the non-interacting
Hamiltonian
\begin{equation}
  H_0 = \sum_{ia\alpha} \int_{\bf k}^\Lambda \! \left\{ \epsilon_{ia}({\bf k})
    - \mu \right\} \psi^\dagger({\bf k}) \psi^{\vphantom\dagger}({\bf
    k}).
  \label{effmassham}
\end{equation}
Here 
\begin{equation}
  \epsilon_{ia}({\bf k}) = (-1)^i \left[ {k_a^2 \over
      {2m^{i}_\parallel}} + {(k^2 - k_a^2) \over {2m^{i}_\perp}} -
    E_G\right].
  \label{dispersion}
\end{equation}
In the Bloch basis, $H_C = \overline{H}_C + H'$, where 
\begin{equation}
  \overline{H}_C = \int_{\bf pqq'}^\Lambda  {2\pi e^2 \over
    {p^2}} \psi^\dagger({\bf
    q+p})\psi^{\vphantom\dagger}({\bf q}) \psi^\dagger({\bf
    q'-p})\psi^{\vphantom\dagger}({\bf q'}),
\end{equation}
describing the long-range part of the Coulomb interaction, and
\begin{equation}
  H'  \!\! = \!\! \int_{\bf pqq'}^\Lambda \!\!\!\! G_{ijkl}^{abcd} 
  \psi_{ia\alpha}^\dagger({\bf p\!+\! q})
  \psi_{jb\alpha}^{\vphantom\dagger}({\bf p}) 
  \psi_{kc\beta}^\dagger({\bf p'\!-\! q})
  \psi_{ld\beta}^{\vphantom\dagger}({\bf p'}).
  \label{Gdef}
\end{equation}
In the $\Gamma \ll 1$ limit, one finds that $G$ can be approximated as 
momentum-independent and $O(e^2/Q^2)$, where 
$Q=O(1)$ is a reciprocal lattice vector.  Thus $H' \ll
\overline{H}_C = O(e^2/q^2)$, and can be regarded as a small
perturbation.   

The approximate ``continuum'' Hamiltonian $H_c = H_0+\overline{H}_C$
has much higher symmetry than the complete one.  Most obviously,
$H_c$ conserves separately the charge and spin of the electrons 
of each flavor, which constitutes a $\left[ U(2)\right]^3$
invariance.  Note that $\overline{H}_C$ has a much larger $U(12)$
symmetry, which is however strongly broken by the inequivalent
dispersion relations in $H_0$.  As emphasized by Halperin and
Rice\cite{HalperinRice}, one may visualize the system at this level as
a collection of (flavored) positrons and electrons, which
naturally have a strong tendency to bind into ``atoms'' -- excitons.
It is thus natural to formally define
an ``s-wave'' excitonic-insulating state by the presence of an
off-diagonal expectation value 
$\langle
\psi^\dagger_{1a\alpha}({\bf r})\psi^{\vphantom\dagger}_{2b\beta}({\bf
  r})\rangle \neq 0$ for some $a,b,\alpha,\beta$.  Clearly, for this
order parameter to 
represent a true symmetry-breaking, there must be no band mixing (at
the points ${\bf Q}_a$) in the kinetic energy.  Mixing is prevented
for the hexaborides as the conduction and valence states at $X$ belong 
to distinct ($X_3^{\vphantom\prime}$,$X_3^\prime$) representations of
the group of the wavevector.  To proceed,
we define local pair fields:
\begin{eqnarray}
  \Psi_{A s}^{\alpha \nu}({\bf r}) & = & {1 \over 2}\psi^\dagger ({\bf r})
  T^{A\alpha} \tau^\nu
  \psi^{\vphantom\dagger}({\bf r}), \\
  \vec{\Psi}_{A t}^{\alpha \nu}({\bf r}) & = & {1 \over 2}\psi^\dagger({\bf r})
  T^{A\alpha} \tau^\nu
  \vec{\sigma}  \psi^{\vphantom\dagger}({\bf r}). 
\end{eqnarray}
Here and in the rest of the paper we introduce Pauli matrices
$\vec{\tau},\vec{\sigma}$ acting in the band and spin
spaces, respectively, and $3\times 3$ matrices $T^{A\alpha}$
($\alpha = 1\ldots n_A$) acting in the ``flavor'' space.  We
also suppress indices wherever possible.  The excitonic order
parameters are off-diagonal in the band space, and so utilize only the
off-diagonal Pauli matrices with $\nu=1,2$.  The
appropriate $T$'s (which specify irreducible representations of the
cubic group -- $\Gamma_1^\prime, \Gamma_3^\prime, \Gamma_5^\prime,
\Gamma_4^\prime$ for $A=1,2,3,4$) are U(3) generators.  In particular,
$T^{11}= I/\sqrt{3}$ and the remainder are expressed in terms of
the standard Gell-Mann basis $\{\lambda_1\cdots\lambda_8\}$:
$T^{21}=\lambda_3/\sqrt{2}$, 
$T^{22} = \lambda_8/\sqrt{2}$, $T^{31} = \lambda_1/\sqrt{2}$, $T^{32}
= \lambda_4/\sqrt{2}$, $T^{33} = \lambda_4/\sqrt{2}$,
$T^{41}=\lambda_2/\sqrt{2}$, $T^{42} = \lambda_5/\sqrt{2}$, $T^{43} =
\lambda_7/\sqrt{2}$. 
These satisfy ${\rm Tr} \, T^{A\alpha} T^{B\beta} =
\delta^{AB}\delta^{\alpha\beta}$.  
Due to the separate $U(2)$ invariances for each $i a$ in $\psi_{ia\alpha}$,
the order parameters are unified for $H_c$ into only two independent
multiplets.  The first contains the diagonal (in flavor) order
parameters with $\Psi_{A s}^{\alpha \nu}, \vec{\Psi}_{A
  t}^{\alpha \nu}$, with $A=1,2$; the second contains the remaining
off-diagonal order parameters with $A=3,4$.  

The energetics separation of the above two multiplets is entirely due
to the anisotropy in dispersion at the different ${\bf Q}_a$ points.
Several points suggest that this anisotropy 
generally favors the {\sl diagonal} order parameters.  A somewhat physical
argument is that electrons and holes that can best move together tend to
pair more effectively.  This suggests that states with similar group
velocity at equal momenta (necessary for a zero-momentum condensate)
and hence similar dispersion will tend to preferentially bind.  Note
that anistropies in the interactions also exist, but are much weaker
than those in the dispersion.

These heuristic considerations are supported by two concrete
calculations.  Firstly, consider approach the EI state by reducing
$E_G>0$ in the band insulator.  In this limit, the nature of the
condensate is determined by the lowest-energy bound state (which
reaches the chemical potential first).  It is then necessary to
compare the binding energies of electron-hole pairs taken from the
same and different ${\bf K}_a$.  As a caricature of the true band
structure, we consider a toy model with $m_{\parallel,\perp}^1 =
m_{\parallel,\perp}^2 \equiv m_{\parallel,\perp}$ (c.f.
Eq.~\ref{dispersion}).  For an electron-hole pair drawn from the same
X point, the effective masses match, and the reduced mass in the
corresponding anisotropic hydrogen atom problem is $m_r =
(m_\parallel/2,m_\perp/2,m_\perp/2)$.  If the electron and hole are
taken from different X points, one obtains instead $m_r =
(m_\perp/2,\tilde{m},\tilde{m})$, with $\tilde{m} = m_\parallel
m_\perp/(m_\parallel + m_\perp)$.  Various methods (e.g. variational)
can be used to 
convince oneself that the ``diagonal'' electron-hole pair is more
strongly bound whenever $m_\parallel \neq m_\perp$.  This is particularly
clear in the limit $m_\parallel \gg m_\perp$, in which case the
diagonal atom has one very heavy reduced mass component, while all the
reduced masses become equal to $m_\perp/2$ in the off-diagonal case.

A second argument is provided by diagrammatic techniques which apply
deep into the metallic limit when $E_G <0$ and Coulomb effects are
{\sl weak}, $e^2\sqrt{m}/E_G\ll 1$.   It is then
possible to integrate out modes and further reduce the momentum-space
cut-off to a set of spherical shells of width $2\Lambda' \ll k_F$
around the Fermi surfaces\cite{CoulombNote}.  With the
reduced cut-off, fermi-surface renormalization group (RG)
arguments\cite{Shankar}\ or more traditional diagrammatics may be used
to show that the only (marginally) relevant two-particle interactions
(i.e. which lead to logarithmic instabilities) are
valence-conduction ``ladder'' terms in which a conduction electron is
scattered to a valence state at a nearby point on the Fermi surface
and vice versa.
The most general allowed terms of this type can be written
\begin{equation}
  H_C  =  -{1 \over 2}\sum_{\nu A\alpha} \int^{\Lambda'} \Bigg[
  V^{\nu A\rho} \left|\Psi_{As}^{\alpha \nu}\right|^2 
 + V^{\nu A\sigma} \left|\vec{\Psi}_{At}^{\alpha\nu}\right|^2\Bigg],
  \label{irrepexpansion}
\end{equation}
\noindent where the superscript indicates the spherical shell cut-off,
and $\nu=1,2$ sums only over the inter-band 
$\vec{\tau}$ matrices.  Furthermore, to simplify the presentation we have
kept only the most relevant s-wave component of each interaction
channel; including the full wave-vector dependence is straightforward
and leads to no significant modifications.  Note that
with the sign convention above $V>0$ is an effectively ``attractive''
interaction.  Due to its high symmetry, the continuum interaction
$\overline{H}_C$ gives {\sl identical} contributions
to all the coupling constants, so $V^{\nu A \rho/\sigma} =
V_0 + \delta V^{\mu A \rho/\sigma}$, with $V_0 \approx 2\pi
e^2\ln(k_F/\Lambda')/k_F^2$ and
\begin{eqnarray}
  \delta V^{\nu A \rho} & = & \left(G_{ilkj}^{adcb}-2
    G_{ijkl}^{abcd}\right) \tau_{ji}^\nu 
  \tau_{lk}^\nu T_{ba}^{A\alpha} T_{dc}^{A\alpha}, \\
  \delta V^{\nu A \sigma} & = & G_{ilkj}^{adcb}\tau_{ji}^\nu
  \tau_{lk}^\nu T_{ba}^{A\alpha} T_{dc}^{A\alpha}. 
\end{eqnarray}
Note that the corrections to the singlet interactions, $\delta V^{\nu
  A\rho}$, have a repulsive ``direct'' contribution absent in the
triplet ($\sigma$) channel, so that the correction terms favor triplet
pairing, in agreement with the Hartree-Fock theory of
Ref.~\onlinecite{HalperinRice}.  Amongst the favored flavor-diagonal
representations, there are four possible triplet order parameters,
with $A=1,2$, $\mu=1,2$.  To decide between them requires an
explicit calculation of the Coulomb matrix elements, and hence
knowledge of the Bloch states.  The simplest wavefunctions consistent
with the $X_3$,$X_3^\prime$ symmetries are $\phi_1 = \sqrt{8}\cos \pi
x \sin 2\pi y \sin 2\pi z$, $\phi_2 = \sqrt{2}\sin \pi x (\cos 2\pi y
- \cos 2\pi z)$ for the valence and conduction states at $(\pi,0,0)$,
respectively.  For these wavefunctions, one finds the most favorable
correction occurs to the $A=2$,$\nu=1$ coupling constant, suggesting
an EI in the $\Gamma_3^\prime$ representation.  The presence of
$\tau^x$ pairing indicates broken time-reversal symmetry consistent
with a spatially-varying spin density within the unit cell.  This
result should not be taken as definitive, and might be modified by
e.g. including higher harmonics in the Bloch functions.  It is crucial
to remember, however, that the splittings due to the $\delta V$ terms
are only weak perturbations to the continuum contribution from
$\overline{H}_C$.  Thus although a particular EI state is
energetically preferred, the other nearly degenerate states can play
important roles.

Consider next a mean-field (MF) approach\cite{Volkov,Zhito}. Following
the above reasoning, we include only the diagonal interactions,
$V^{\nu A \rho}, V^{\nu A \sigma}$, with $\nu =1,2$, $A=1,2$.  To
allow analytic progress, we also assume the toy model dispersion
above and collinear spin polarization of all non-zero order parameters
along the $\sigma^z$ axis.  Define gap functions $\Delta_{a\alpha}^\nu
= {V_0 \over 2}\langle\psi^\dagger_{a\alpha}\tau^\nu
\psi^{\vphantom\dagger}_{a\alpha}\rangle \equiv (\Delta_{a s}^\nu +
\alpha \Delta_{a t}^{\nu,z})/2$.  The mean-field hamiltonian
neglecting $\delta V$ corrections is $H_{MF} = \int_{\bf k}^{\Lambda'}
{\cal H}_{MF}({\bf k})$ with,
\begin{equation}
  {\cal H}_{MF} = \sum_{\mu a\alpha} \left\{
    \psi^\dagger_{a\alpha}  \left[ \epsilon^a_{\bf k} \tau^z -
      \mu - \Delta_{a\alpha}^\mu \tau^\mu \right]\psi_{a\alpha} + {1
      \over {V_0}}
    \left[\Delta_{a\alpha}^\mu\right]^2 \right\}. 
  \label{mfham}
\end{equation}
Because ${\cal H}_{MF}$ is a sum of independent terms for different
flavors and spin orientations, each such species can be studied
separately.  To proceed, we note that the hamiltonian for a single
species can be mapped to the problem of a BCS superconductor in a
Zeeman field.  Under the particle/hole transformation
$\psi_{2a\alpha}^{\vphantom\dagger} \rightarrow
\psi_{2a\alpha}^\dagger$, $\Delta_{a\alpha}^1 + i \Delta_{a\alpha}^2$
becomes the complex BCS order parameter and $\mu$ the Zeeman field, if
valence and conduction states are re-interpreted as up and down spins.
This MF theory (MFT) was solved by Larkin and
Ovchinnikov\cite{Larkin}\ and Fulde and Ferrell\cite{Fulde}.  For
small $\mu$ at low temperatures, a BCS state occurs, with
$|\Delta|=\Delta_0 = 2t \exp(-1/N_0 V_0)$ at $T=0$, where $t=v_F
\Lambda'$ is an 
energy cut-off and $N_0$ is the density of states at
the Fermi level.  In the simplest MFT assuming a uniform order
parameter, at zero temperature a first order transition occurs at
$\mu=\mu_c=\Delta_0/\sqrt{2}$ directly into the normal state.  A more
detailed examination shows that an intermediate state with a
non-uniform order parameter\cite{Larkin,Fulde}\ exists in a narrow
region around 
$\mu=\mu_c$, with a first order transition to the BCS state at
$\mu=\mu_c'<\mu_c$ and a possibly second order transition to the
normal state at $\mu=\mu_c''>\mu_c$.  Note that the ``non-trivial''
solution used in Refs.~\onlinecite{Zhito,Volkov}\ is actually an
unstable and unphysical free energy {\sl maximum}.

For our purposes, it is sufficient at this stage to consider only the
uniform MFT -- in any case the non-uniform solution is invalidated by
the correction terms soon to be added to Eq.~\ref{mfham}.  The above
analysis indicates that for $\mu=\mu_c$, there are two degenerate
minima of the single species free energy density
\begin{equation}
  f_{a\alpha} = - k_{\rm B}T\int_{\bf k}^{\Lambda'} \sum_\pm \ln [1
  + \exp(-\beta \xi_\pm({\bf k}))] +
  {\left[\Delta_{a\alpha}^\mu\right]^2 \over V_0}.
\end{equation}
Here $\beta = 1/(k_B T)$ and the quasiparticle energies are $\xi_\pm =
-\mu \pm \sqrt{\epsilon_{\bf k}^2 + \Delta^\mu \Delta^\mu}$.
Since the total free energy density is the sum $f_{\rm tot} =
\sum_{a\alpha} f_{a\alpha}$, this implies a very large ($2^{2N}$-fold)
degeneracy for the full electron-hole system.  In particular, the amplitude
for each flavor and spin orientation may be chosen independently to
equal $0$ or $\Delta_0$.   While two equal free energy minima
occur at any first order transition, this large degeneracy is
non-generic.


Additional interactions can split the degeneracy in favor of a
ferromagnetic state.  The search for symmetry-breaking terms is
simplified by the fact that, because the different states have
macroscopically different particle occupations, only operators
diagonal in flavor and spin have 
non-vanishing matrix elements in the degenerate subspace.  A little
examination shows that the dominant perturbations are 
\begin{equation}
  H^{s} \! = \!\! -\!{1 \over
    2}\!\sum_{\stackrel{A=1,2}{\alpha}}\!\int^{\Lambda'}\! 
  \Bigg[  
  \delta_{A2}W^{2\rho} \left|\psi^\dagger
    T^{2\alpha}\psi^{\vphantom\dagger}\right|^2 
 \!\!+\! W^{A\sigma} \!\left| \psi^\dagger
   T^{A\alpha}\vec{\sigma}  \psi^{\vphantom\dagger} \right|^2
  \Bigg]. 
  \label{splits}
\end{equation}
A $W^{1\rho}$ term, which represents the
long-wavelength Coulomb potential, was neglected, as appropriate for
uniform states provided the background charge is taken 
into account (but see below).  
Terms involving
$\psi^\dagger\tau^z\psi^{\vphantom\dagger}$  were omitted from
Eq.~\ref{splits}, as they are important only for band-gap
renormalization.  Likewise, the
$\delta V$ corrections directly couple the $\Delta_{a\alpha}$ order
parameters, but are negligibly small: $\delta V \ll
W^{A\rho/\sigma} \approx V_0$.

Treating the terms in Eq.~\ref{splits}\ as perturbations, one finds
that to the leading order approximation in which
$W^{1\rho}=W^{1\sigma}=W^{2\sigma}$, the lowest energy states for the
cubic problem comprise two sets: $8$ with ferromagnetic (but but
arbitrarily oriented) polarizations of all flavors (e.g. $\Delta_{a+}
= \Delta_0, \Delta_{a-}=0$), and $12$ with one X-point ungapped, one
X-point polarized, and one fully gapped (e.g.
$\Delta_{1\pm}=0,\Delta_{2+}=\Delta_0,\Delta_{2-}=0,\Delta_{3\pm}=\Delta_0$).
The next order corrections ($\delta W^{1\rho}<\delta W^{2\sigma}<0,
\delta W^{1\sigma}>0$) favor the aligned and fully polarized state
(note that transverse magnetization $\vec{M} \wedge
\vec{\Delta}^{\alpha\mu}_{At} \neq 0$ is never favored).

To understand the behavior with chemical potential in more detail, we
expand the MFT and introduce order parameters decoupling 
the $W^{1\sigma}$ and $W^{2\sigma}$ spin interactions (the $W^{1\rho}$ 
interaction can also be decoupled, but does not influence the
mean-field solution in the physical parameter regime).  The extended
MFT is obtained by replacing $\mu \rightarrow \mu
+ \alpha h_a$ for flavor,spin $a,\alpha$ and adding the terms
\begin{equation}
  {\cal H}_h = {1 \over
      {2J}}(h_1^2\!+\! h_2^2\!+\! h_3^2)\! -\! {\delta \over J^2}(h_1
    h_2\!+\! h_2 h_3 \!+\! h_1 h_3) 
  \label{hham}
\end{equation}
to the original hamiltonian density ${\cal H}_{MF}$.  Here $J =
\langle 2\pi e^2/ K^2\rangle \gg \delta \sim
e^2/ Q^2 > 0$.  It is straightforward to solve the
extended MFT for $\delta \ll J$ at zero temperature.  One finds a
ferromagnetic ground state in the range $\mu_{c1} = \mu_c\sqrt{1-2N_0
  J}<\mu<\mu_{c2} = \mu_c\sqrt{(1-2N_0J)/(1-4N_0 J)}$, with first
order transitions at 
$\mu_{c1},\mu_{c2}$ (this assumes $\mu_{c2}<\Delta_0$).  At finite 
temperature the MFT gives the phase diagram shown in Fig.~1.  Note
that the above behavior is stable to further perturbations such as
the $\delta V$ corrections, except insofar as to stabilize a
particular (e.g. $\Gamma_3^\prime$) triplet state for $\mu<\mu_{c1}$.

A salient feature of the MFT is the first-order nature of the
low-temperature boundary of the ferromagnetic phase.  
Taking into account long-range Coulomb interactions, macroscopic charge
neutrality requires that {\sl as a function of doping $x$} the system must
therefore form an inhomogeneous (i.e. domain or labyrinthine) state for
$0<x<x_c = 6N_0 \mu_{c1}$, the minimum value in the ferromagnetic
phase.  It is straightforward to estimate the typical 
width $\ell$ of ferromagnetic domains (their detailed morphology is
much more difficult to determine) which is controlled by a competition
between electrostatic energy and surface tension $\sigma \sim N_0 v_F
\Delta_0$, which set the profile
of the electrochemical potential.  One finds $\ell \sim \left[
  \sigma/(e^2 x^2 (1-x/x_c)^2)\right]^{1/3}$.  Note that if $\ell$
becomes of order the 
``coherence'' length $\xi = v_F/\Delta_0$, this semiclassical analysis 
is invalid and a microscopic quantum calculation is required.  A
consequence of the domain state is that the magnetization density is
{\sl linear} in the doping in this region, i.e. $m/x$ is constant (and 
stays non-zero as $x\rightarrow 0^+$).  Similar behavior holds for
$x_c'=6N_0\mu_{c2} < x < x_c'' = 12 N_0\mu_{c2}$, where a non-uniform mixture
of ferromagnetic and normal domains obtains.  

Beyond the simplest MFT, it is possible for the system to remain in
an unpolarized and uniform excitonic state for very small doping at zero
temperature.   This can occur only if the undoped excitonic state is
{\sl itinerant} (and therefore gapless).  To study the
feasibility of this scenario, we have considered an {\sl unnested} MFT
(i.e. with different effective masses for electrons and holes).
In this case, an intermediate-coupling {\sl excitonic metal} phase
occurs near $\mu=0$, with a non-vanishing excitonic order
parameter but an incompletely gapped Fermi surface.  A first order $T=0$
transition to an excitonic ferromagnet remains if the
incommensurability is not too large.  It is likely that
a continuous excitonic paramagnet--ferromagnet transition is
prohibited by the singular quasiparticle-mediated interactions, but
this issue warrants further investigation.  

Many open questions remain.  An enormous multitude of collective
(pseudo-Goldstone modes of $[U(2)]^3$) modes
should exist in the $|\Gamma| \ll 1$ limit, mostly with small gaps.
There are also a large number of metastable free-energy minima
which could lead to substantial hysteresis.  Non-trivial topological
excitations are also possible.  Finally, the role of disorder and
possible trapping by dopant ions should be considered.  The present
analysis provides a framework for systematic studies of these issues,
and of the applicability of the excitonic model to the hexaborides.

\acknowledgements

We thank Z. Fisk, H. R. Ott, and L. Gorkov for discussions.


\vskip -0.2in

\begin{references}

\bibitem{Hexaborides} D. P. Young {\sl et. al.}, Nature {\bf 397}, 412 
  (1999).

\bibitem{HalperinRice} B. I. Halperin and T. M. Rice, in {\sl Solid
    State Physics}, {\bf 21}, F. Seitz, D. Turnbull, and
  H. Ehrenreich, eds., Academic Press, New York, 1968.

\bibitem{Monnier} S. Massidda {\sl et. al.}, Z. Phys. B {\bf 102}, 83
  (1997); A. Hasegawa and A. Yanase, J. Phys. C {\bf 12}, 5431 (1979).

\bibitem{Zhito} M. E. Zhitomirsky, T. M. Rice and V. I. Anisimov,
  cond-mat/9904330, unpublished. 

\bibitem{Volkov} B. A. Volkov {\sl et. al.}, Sov. Phys.-JETP {\bf 41}, 
  952 (1976); {\sl ibid.}, {\bf 43}, 589 (1976).

\bibitem{Larkin} A. I. Larkin and Yu. N. Ovchinnikov, Sov. Phys.-JETP
  {\bf 20}, 762 (1965).

\bibitem{Fulde} P. Fulde and R. A. Ferrell, Phys. Rev. {\bf 135}, A550 
  (1964).

\bibitem{inprep} L. Balents, in preparation.


\bibitem{CoulombNote} A proper treatment of the long-range Coulomb
  interaction in the RG approach will be deferred to Ref.~\onlinecite{inprep}.

\bibitem{Shankar} R. Shankar, Rev. Mod. Phys. {\bf 66},  129  (1994).



\end{references}
\end{document}